\documentclass[preprint]{aastex631}

\usepackage{changepage}

\begin{document}

\title{The Coldest Known Y Dwarfs: Estimates of their Effective Temperatures}

\correspondingauthor{Sandy Leggett}
\email{sandy.leggett@noirlab.edu}

\author[0000-0002-3681-2989]{S. K. Leggett}
\affiliation{Gemini Observatory/NSF's NOIRLab, 670 N. A'ohoku Place, Hilo, HI 96720, USA}


\begin{abstract}
For a decade there has been a factor of 2.5 gap in luminosity between the 275~K WISE J085510.83-071442.5 \citep{Luhman_2014} and all other Y dwarfs, with $T_{\rm eff} \gtrsim 350$~K.  Recently three objects were found which may fall in this gap.   
Two are companions to Y dwarfs: WISE J033605.05-014350.4B \citep{Calissendorff_2023} and CWISEP J193518.58-154620.3B \citep{Furio_2025}; the third is MEAD 62B, a candidate companion to a white dwarf \citep{Albert_MEAD_2025}.   Evolutionary models 
calculate a tight relationship between  luminosity and $T_{\rm eff}$ for Y dwarfs.  I determine luminosities and hence $T_{\rm eff}$ for three Y dwarfs
(WISE J085510.83-071442.5, WISE J173835.53+273259.0, WISE J182831.08+265037.7). I derive relationships between $T_{\rm eff}$ and mid-infrared colors using these together with 
22 T and Y dwarfs from \citet{Beiler_2024} 
with luminosity-based $T_{\rm eff}$ values.  
These relationships are used to explore the  $T_{\rm eff}$ distribution for Y dwarfs. A sample of 31 Y dwarfs within $\sim 20$~pc is presented with $275 < T_{\rm eff}$~K $< 425$.   The {\it JWST} colors for WISE J053516.80-750024.9 and WISE J182831.08+265037.7 support previous suggestions that they are unresolved binaries, the former a  480~K and 340~K dwarf pair and the latter a pair of 387~K dwarfs.  Five other dwarfs have unusual colors; two are likely high gravity and/or metal-poor (WISE J024714.52+372523.5, WISEA J215949.54-480855.2), two low gravity and/or metal-rich (CWISEP J104756.81+545741.6, WISE J150115.92-400418.4), and the fifth cannot be interpreted (WISE J043052.92+463331.6).
An Appendix provides colors  which can be used as a reference for searches for brown dwarfs  in  {\it JWST} data.

\end{abstract}



\bigskip
\section{Introduction} 

Brown dwarfs have insufficient mass for sustained energy production by nuclear fusion, and cool over periods of gigayears \citep[e.g.][their Figure 13]{Marley_2021}. The coldest brown dwarfs are classified as Y dwarfs, and have effective temperatures ($T_{\rm eff}$) less than 500~K \citep{Cushing_2011, Kirkpatrick_2021a}. The warmer T-type brown dwarfs have $500 \lesssim T_{\rm eff}$~K $\lesssim 1200$ \citep[e.g.][their Figure 22]{Kirkpatrick_2021a}. Note that spectral type is based on the appearance of near-infrared spectra and not on $T_{\rm eff}$ \citep{Burgasser_2006a,Cushing_2011,Kirkpatrick_2012}.

The discovery of WISE J085510.83$-$071442.5 (hereafter WISE 0855-07) was presented by \citet{Luhman_2014}. It is a remarkably cold brown dwarf, with a $T_{\rm eff}$ of only 275~K \citep{Luhman_2024, Kuhnle_2024, Leggett_2025b}; it is also very close, only 2~pc from the Sun \citep{Luhman_2016}. It holds the record for the least luminous brown dwarf known with log $L/L_{\odot} \approx -7.2$. Until very recently there was a large gap in luminosity between
it and all other known brown dwarfs, with  the next luminous brown dwarf being more than $2.5\times$ brighter, having  $T_{\rm eff} \gtrsim 350$~K and log $L/L_{\odot} \gtrsim -6.8$.   

Scientists using the {\it James Webb Space Telescope} ({\it JWST}) are now discovering objects within this temperature and luminosity gap, thanks to {\it JWST}'s sensitivity to mid-infrared light, and the instruments' high spatial resolution.
Two of the objects that lie in the gap are close companions to other Y dwarfs, forming the tight binary systems WISE J033605.05-014350.4AB \citep[hereafter WISE 0336-01AB]{Calissendorff_2023}
and CWISEP J193518.58-154620.3AB \citep[hereafter WISE 1935-15AB]{Furio_2025}.  A third object is a candidate brown dwarf companion to the white dwarf MEAD 62 \citep[also known as  2MASS J09424023-4637176,][]{Albert_MEAD_2025}. The analyses by \citet{Albert_MEAD_2025} confirm that MEAD 62B is an unresolved point source, however they estimate that there is a 50\% chance that the object is a false-positive red dot. Multi-epoch observations are needed to measure a proper motion for the source, to determine whether or not it is associated with the white dwarf. The mid-infrared colors and luminosity of the source are consistent with it being a cold brown dwarf at the distance of the white dwarf.

In this work I explore the relationships between mid-infrared luminosity, color, and $T_{\rm eff}$ for a sample of brown dwarfs, with particular emphasis on the coldest Y dwarfs. 
Using a sample of brown dwarfs with $T_{\rm eff}$ values determined from measurements of bolometric luminosity, relationships are determined which can be used to estimate $T_{\rm eff}$ for other objects. 

In Section 2, I calculate the luminosities and temperatures for three Y dwarf systems, and update the values for a T dwarf system. In Section 3, I use these values, together with those determined by \citet{Beiler_2024} for a sample of T and Y dwarfs, to produce regression fits to  color-$T_{\rm eff}$ datasets.  Brown dwarfs which deviate from the general population are discussed.   In Section 4 the population of the coldest Y dwarfs is explored. In Section 5 the physical parameters of the three binary systems with cold Y dwarf secondaries are described. My conclusions are given in Section 6.
The Appendix 
presents a reference set of
{\it JWST} colors which are useful 
for deep searches for cold brown dwarfs in {\it JWST} data, such as the JADES and CEERS surveys \citep{Hainline_2024, Hainline_2025}.

\bigskip
\section{Luminosities and Effective Temperatures for Y Dwarfs}

The theory of how brown dwarfs evolve, in terms of their mass, radius and luminosity, 
is well established \citep[e.g.][]{Burrows_1993, Burrows_1997}. Comparison of the evolutionary models to measurements of radius and mass for brown dwarfs and giant planets shows good agreement \citep{Marley_2021}.  Figure 1 is a luminosity:$T_{\rm eff}$ plot, with data taken from the evolutionary models of \citet{Marley_2021}.  
Isochrones are shown for ages of 0.5~Gyr and 10~Gyr; these ages span the likely range in age for a local sample of brown dwarfs  \citep[e.g.][]{Buder_2019}, although the population synthesis of brown dwarfs within 25~pc by \citet{Best_2024} shows that more than 90\% of the population is younger that 4~Gyr.  As luminosity decreases the range in 
$T_{\rm eff}$ becomes small, so that luminosity significantly constrains $T_{\rm eff}$ for cold brown dwarfs. Note that mass is not constrained by luminosity --- in order to estimate mass a measurement of surface gravity is required \citep[see][their Figure 10]{Marley_2021}.

\begin{figure}
\vskip -1.2in
\hskip 0.3in
\includegraphics[angle=-90, width = 6.5 in]
{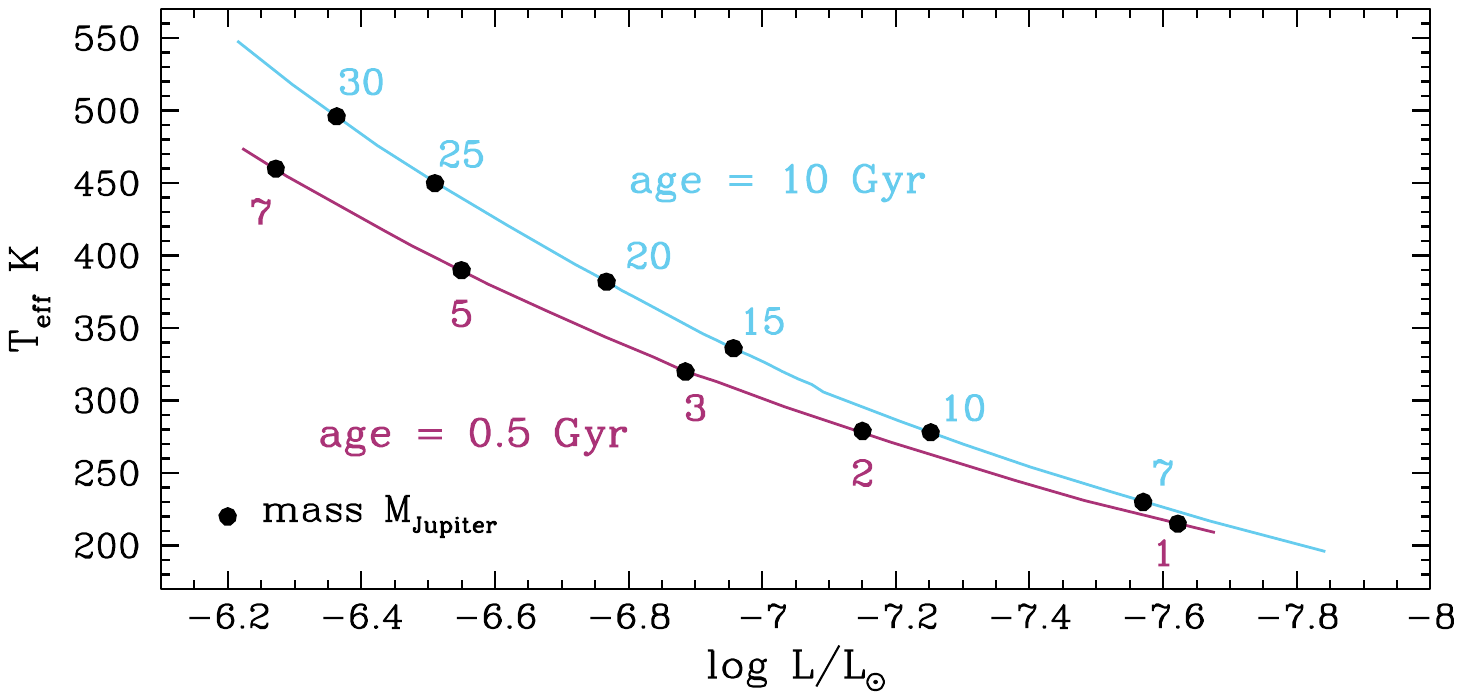}
\vskip -0.2in
\caption{Relationship between effective temperature and luminosity for solar-metallicity brown dwarfs, from the evolutionary models by \citet{Marley_2021}. The upper blue line is for older brown dwarfs with an age of 10~Gyr, and the lower red line is for younger brown dwarfs with an age of 0.5~Gyr. Dots along the sequences indicate mass in units of $M_{\rm Jupiter}$. 
}
\end{figure}

In this work I use a sample of brown dwarfs that have published spectra covering a broad range of wavelength including the mid-infrared, where most of the energy is emitted by cold objects.  With these data, a reliable determination of bolometric luminosity can be made, constraining  $T_{\rm eff}$ (Figure 1). 
The \citet{Beiler_2024} sample is the primary data source; low-resolution NIRSpec and MIRI spectra were obtained for twenty-two T and Y dwarfs covering the wavelength range $0.8 \leq \lambda~\mu$m $\leq 12.0$. Longer wavelength photometry was also obtained using MIRI with the F1000W, F1500W, F1800W and F2100W filters. \citet{Beiler_2024} extended the spectral energy distribution (SED) beyond $21~\mu$m by adopting a Rayleigh–Jeans flux distribution.  For one of the T dwarfs in the \citet{Beiler_2024} sample, SDSSp J134646.45-003150.4, \citet{Vrba_2026} determined a parallax that is more precise and significantly different from the value used by \citet{Beiler_2024}. I have used the newer parallax to revise the luminosity and $T_{\rm eff}$ for this object, and the values are given in Table 1. The brown dwarf is $\approx 10\%$ closer, making the luminosity $\approx 20\%$ fainter, and  $T_{\rm eff} \approx 5\%$ lower.

\begin{deluxetable*}{lrrrrr}
\tabletypesize{\footnotesize}
\tablecaption{Luminosity and Implied $T_{\rm eff}$ for an Age Range of 0.5 to 10~Gyr}
\tablehead{
\colhead{} &  \colhead{Distance$^*$} &  \colhead{$F_{\rm bol}$} &
\colhead{$L_{\rm bol}$} &
\colhead{log $L/L_{\odot}$} &
\colhead{$T_{\rm eff}$}\\
\colhead{Name} &  \colhead{pc} &  \colhead{$10^{-16}$ W m$^{-2}$} &
\colhead{$10^{20}$ W} &
\colhead{} &
\colhead{K}
}
\startdata
WISE J085510.83-071442.5 & 2.28 $\pm$ 0.01 & 3.71  $\pm$ 0.15 & 0.230  $\pm$ 0.009 & -7.22 $\pm$ 0.02 & 276 $\pm$ 9 \\
SDSSp J134646.45-003150.4** & 13.20 $^{+0.22}_{-0.06}$  & 11.26 $\pm$ 0.16  & 23.5 $\pm$ 1.6  & -5.21 ± 0.03 & 927 $^{+58}_{-87}$ \\
WISE J173835.53+273259.0 & 7.64 $\pm$ 0.12 &  1.77 $\pm$ 0.14 & 1.24 $\pm$ 0.15 & -6.49 $\pm$ 0.05 & 431 $\pm$ 27 \\ 
WISE J182831.08+265037.7 & 9.97 $\pm$ 0.19 & 1.39 $\pm$ 0.04 & 1.65 $\pm$ 0.04 & -6.37 $\pm$ 0.02  & 387 $\pm$ 22*** \\
\enddata
\smallskip
\tablenotetext{*}{Distances are determined from parallaxes published in \citet{Kirkpatrick_2021a}, except for SDSSp J134646.45-003150.4 for which the parallax published by \citet{Vrba_2026} is used.}
\tablenotetext{**}{~~The luminosity and $T_{\rm eff}$ have been revised from the values determined by \citet{Beiler_2024} using the parallax published by \citet{Vrba_2026}. The uncertainty in  $T_{\rm eff}$ has been set to that quoted by \citet{Beiler_2024}.}
\tablenotetext{***}{~~~The  $T_{\rm eff}$ is determined assuming WISE J182831.08+265037.7 consists of an identical pair of Y dwarfs.  If this object is single then the luminosity implies $T_{\rm eff} = 462 \pm 25$.}
\end{deluxetable*}

I have added three Y dwarfs to the luminosity sample, which all have published MIRI spectra:
\begin{itemize}
    \item WISE 0855-07 --- {\it JWST} NIRSpec spectra covering $0.8 \leq \lambda~\mu$m  $\leq 5.5$ was published by \citet{Luhman_2024}. {\it JWST} MIRI spectra covering $5.2 \leq \lambda~\mu$m  $\leq 22$ was published by \citet{Kuhnle_2024}. I extended the SED beyond $22~\mu$m by adopting a Rayleigh–Jeans flux distribution.  
    The observed portion of the bolometric flux makes up 84\% of the total, with the Rayleigh–Jeans extension making up 16\%. For the observational fraction of the SED, I assume that the uncertainty equals the uncertainty in the 
    \href{https://jwst-docs.stsci.edu/jwst-calibration-status#gsc.tab=0}
{absolute flux calibration of the instruments, which is currently estimated to be 3\%}. Estimating an uncertainty of 10\% for the    Rayleigh–Jeans tail (based on the uncertainty of defining a ``continuum'' point at which to start the extension), the total uncertainty in the bolometric flux is 4\%. 
I adopt a range for  $T_{\rm eff}$ assuming a possible range in age of 0.5~Gyr to 10~Gyr. The results are given in Table 1.
    \item WISE J173835.53+273259.0 (hereafter WISE 1738+27) --- near-infrared ground-based spectra covering $1.0 \leq \lambda~\mu$m  $\leq 2.2$ was published by \citet{Cushing_2011} and \citet{Leggett_2016a}.   {\it JWST} MIRI spectra covering $4.9 \leq \lambda~\mu$m  $\leq 18$ was published by \citet{Vasist_2025}.  I use
    \href{https://www.erc-atmo.eu/?page_id=322}{
    ATMO 2020++ without PH$_3$
    synthetic spectra} \citep{Phillips_2020, Leggett_2021, Meisner_2023, Leggett_2025a} to bridge the $2.2 \leq \lambda~\mu$m  $\leq 4.9$ gap , selecting spectra from the available grids which reproduce the observed spectra and the {\it WISE} W2 and {\it JWST} F480M photometry measured for this dwarf \citep[ALLWISE Catalog and][]{Albert_2025}.  A $T_{\rm eff} = 400$~K, log $g = 4.0$ solar metallicity synthetic spectrum reproduces the data well. This spectrum was also used to extend the SED to $\lambda = 30~\mu$m (where the model terminates), beyond which a Rayleigh–Jeans tail was adopted.  For this object, the observed portion of the SED makes up 63\% of the total flux, the modelled  $2.2 \leq \lambda~\mu$m  $\leq 4.9$ region around 24\%, and the region at $\lambda > 18~\mu$m 13\%. By exploring variations in the the model fluxes around the critical $\lambda \sim 4.5~\mu$m region, I estimate an uncertainty in the total bolometric flux of 8\%.  Again a range in age of 0.5~Gyr to 10~Gyr is adopted, and $T_{\rm eff}$ values determined accordingly. The results are given in Table 1.
    \item WISE J182831.08+265037.7 (hereafter WISE 1828+26) --- {\it JWST} NIRSpec low-resolution spectra covering $0.8 \leq \lambda~\mu$m  $\leq 5.3$ was obtained as part of the {\it JWST} Cycle 1 GTO program 1189, PI: Thomas Roellig. Higher resolution NIRSpec spectra overing $2.9 \leq \lambda~\mu$m  $\leq 5.1$ was
    published by \citet{Lew_2024}. {\it JWST} MIRI spectra covering $4.9 \leq \lambda~\mu$m  $\leq 18$ was published by \citet{Barrado_2023}.  This object has been recognized to be significantly super-luminous, but has not been resolved into a multiple system in high spatial resolution imaging   \citep[e.g.][]{Beichman_2013, Furio_2023}.  Here I adopt the solution found by \citet{Leggett_2025a} in an exploratory fit of ATMO 2020++ grid spectra to the {\it JWST} data: 
    that the system is a pair of similar cold brown dwarfs with $T_{\rm eff}$ between 350~K and 400~K, surface gravities $g$ in the range log $g$ 4.5 to 5.0 dex, with solar or slightly subsolar metallicity. In order to complete the 
  SED for this work, I  used two $T_{\rm eff} = 400$~K, log $g = 5.0$ solar metallicity synthetic spectra to extend the MIRI observations from 18~$\mu$m to 30~$\mu$m, beyond which a Rayleigh–Jeans tail was adopted.  For this object, the observed portion of the SED makes up 91\% of the total, with the model and Rayleigh–Jeans extension making up 9\%. The uncertainty in the bolometric flux is dominated by the  uncertainty in the 
    \href{https://jwst-docs.stsci.edu/jwst-calibration-status#gsc.tab=0}
{absolute flux calibration of the instruments, which is currently estimated to be 3\%}. 
Again a likely range in age of 0.5~Gyr to 10~Gyr is adopted, and $T_{\rm eff}$ values determined accordingly. The results are given in Table 1.
\end{itemize}

\bigskip
\section{Color and $T_{\rm eff}$ for Brown Dwarfs}

\subsection{Dataset}

In this Section I use the luminosity-based $T_{\rm eff}$ values for the twenty two T and Y dwarfs from the \cite{Beiler_2024} sample, together with the three additional Y dwarfs described above, to determine relationships between various colors and $T_{\rm eff}$.  The sample is representative of the local field population, but does not include more extreme objects like  subdwarfs ---  \citet{Beiler_2024} chose targets with a range in color at a given $\lambda \approx 5~\mu$m luminosity \citep[their Figure 1]{Beiler_2024}, but excluded dwarfs suspected to be extremely metal-poor. 

Figures 2 through 4 show $T_{\rm eff}$ as a function of various colors, for the sample of 25 T and Y dwarfs.  I show the relationships for $J -$ W2, $J -$ [4.5], and [3.6] $-$ [4.5] in Figure 2; these are commonly available and frequently used colors for brown dwarfs.  Figure 3 shows the relationships for absolute magnitudes at $\lambda \sim 4.5~\mu$m, where there is a peak in the SED \citep[see e.g.][their Figures 5 -- 7]{Beiler_2024}. The absolute W2 and [4.5] magnitudes are shown, as well as magnitudes for two {\it JWST} filters that have been used in studies of brown dwarfs, F444W and F480M.  F444W and F480M magnitudes were synthesized by  \citet{Beiler_2024} from their data, and I use those values here.
For WISE 0855-07, WISE 1738+27, and WISE 1828+26, the F480M magnitudes are taken from \citet{Albert_2025}.  F444W magnitudes were synthesized for WISE 0855-07 and WISE 1828+26 from the NIRSpec data published by \citet{Luhman_2024} and \citet{Lew_2024}.    Figure 4 shows the relationships for three longer wavelength {\it JWST} filters: F1000W, F1500W, and F1800W.  For WISE 0855-07, WISE 1738+27, and WISE 1828+26, these magnitudes were synthesized 
from the MIRI spectra published by \citet{Kuhnle_2024, Vasist_2025, Barrado_2023}.  The data only span the F1800W filter for WISE 0855-07. Table 2 gives the values of the magnitudes synthesized here.

I do not include the \citet{Beiler_2024} F2100W results because 
\href{https://jwst-docs.stsci.edu/jwst-mid-infrared-instrument#JWSTMidInfraredInstrument-Sensitivityandperformance}
{MIRI's sensitivity} decreases rapidly at wavelengths longer than $18~\mu$m and the F2100W filter is unlikely to be used for cold brown dwarf science.  Also, although some brown dwarf data have been taken using the F1280W filter, the \citet{Beiler_2024} spectra do not extend to wavelengths long enough to span the filter bandpass, and so it cannot be synthesized and included in this work. 

\begin{deluxetable*}{crrrr}
\tablecaption{Synthesized Photometry, Vega Magnitudes}
\tablehead{
\colhead{Name} &  \colhead{F444W} &  \colhead{F1000W} &  \colhead{F1500W} &  \colhead{F1800W} 
}
\startdata
WISE J085510.83-071442.5 & 14.08 & 12.12 & 10.40 & 10.02 \\
WISE J173835.53+273259.0 & \nodata & 12.66 & 11.71 & \nodata \\
WISE J182831.08+265037.7 & 14.45 & 13.11 & 12.02 & \nodata \\
\enddata
\smallskip
\tablecomments{The uncertainty is dominated by the uncertainty in the absolute flux calibration of the {\it JWST} instruments, which is currently estimated to be 3\% or 0.03 magnitudes.}
\end{deluxetable*}

\begin{figure}[b]
\vskip -1.5in
\hskip 0.2 in
\includegraphics[angle=-90, width = 7 in]
{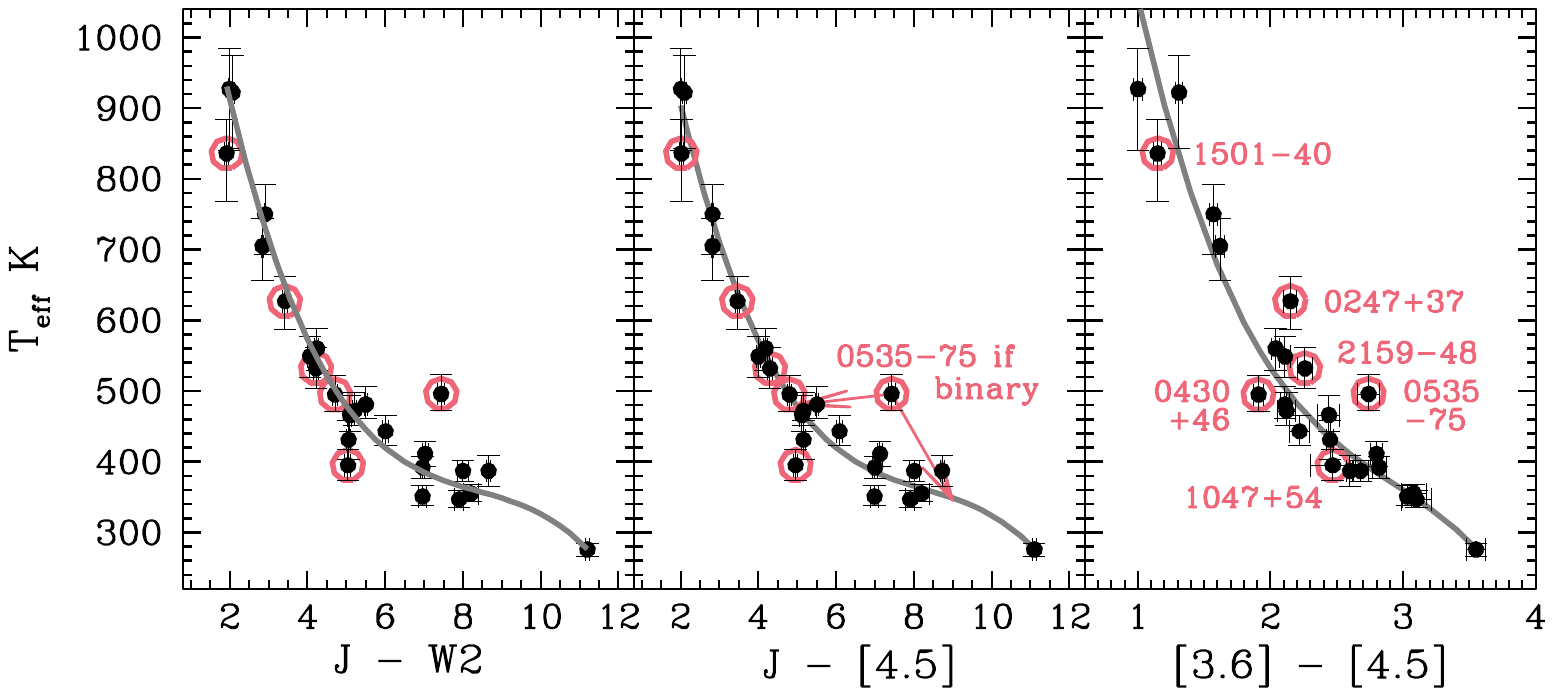}
\vskip -0.3in
\caption{Luminosity-based $T_{\rm eff}$ for the sample of 25 T and Y dwarfs described in Section 2, as a function of $J -$ W2, $J -$ [4.5], and [3.6] $-$ [4.5].  Points circled in red are removed from the regression fits, the results of which are shown as thick gray lines.  Parameters for the fits are given in Table 3. The six brown dwarfs omitted from the fits are identified in the right panel by an abbreviated RA and Decl. WISE 0535-75 may be a binary, composed of a 480~K and 340~K dwarf, as indicated in the middle panel. }
\end{figure}

\begin{figure}
\vskip -0.4in
\hskip 0.5 in
\includegraphics[angle=0, width = 5 in]
{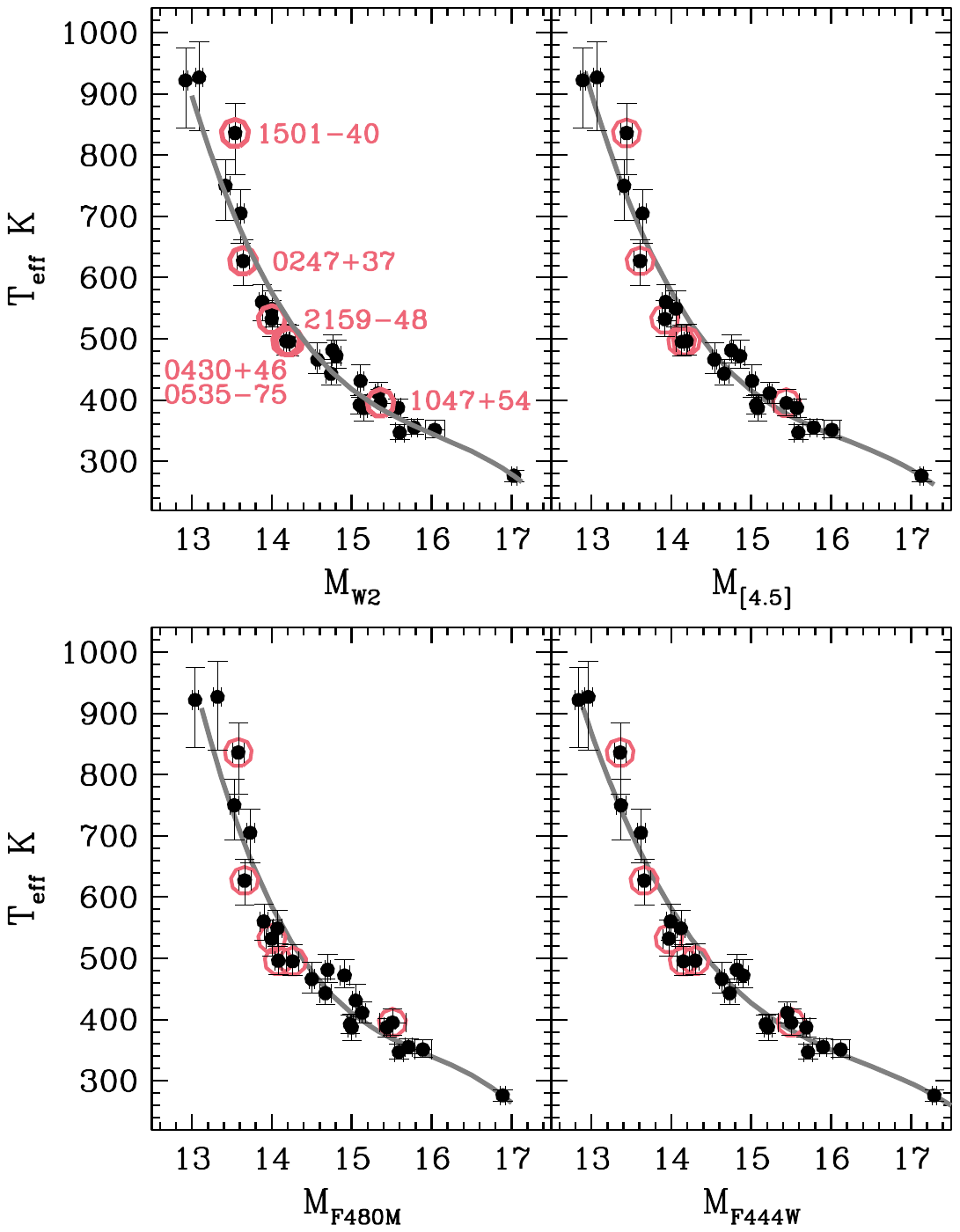}
\vskip -0.5in
\caption{Luminosity-based $T_{\rm eff}$ for the sample of 25 T and Y dwarfs described in Section 2, as a function of absolute magnitudes at $\lambda \approx 4.5~\mu$m.  Six brown dwarfs are omitted from the regression fits, identified in the top left panel, and these points are circled in red.   The results of the fits are shown as gray lines and the parameters are given in Table 3.}
\end{figure}

\subsection{Weighted Regression Fits}

The initial sample for determining a relationship between $T_{\rm eff}$ and color consisted of 25 T and Y dwarfs. Figures 2, 3 and 4 show that there are some obvious outliers in the sequences.  I removed six sources which deviated from the fits by $3\sigma$ or more.  These sources were removed from all fits, including those colors where they did not appear to deviate, in order to have a sample 19 dwarfs with consistent $T_{\rm eff}$ across all colors used here.   (For F444W the sample size is 18, and for F1800W it is 17, because of the incomplete wavelength coverage for synthesizing the F444W and F1800W magnitudes (Table 2).)  Non-linear regression fitting was performed, weighting the $T_{\rm eff}$ values by the uncertainties provided by \citet{Beiler_2024} and Table 1.  Cubic polynomials produced fits with an $rms$ dispersion of $1.2\sigma$ or less, except at F1500W where the $rms$ is $1.3\sigma$ and F1800W where it is $1.8\sigma$. Increasing the number of parameters for the fits did not significantly improve the $rms$.  Table 3 gives the parameters for these fits.

\begin{figure}
\vskip -1.5in
\hskip 0.2 in
\includegraphics[angle=-90, width = 7 in]
{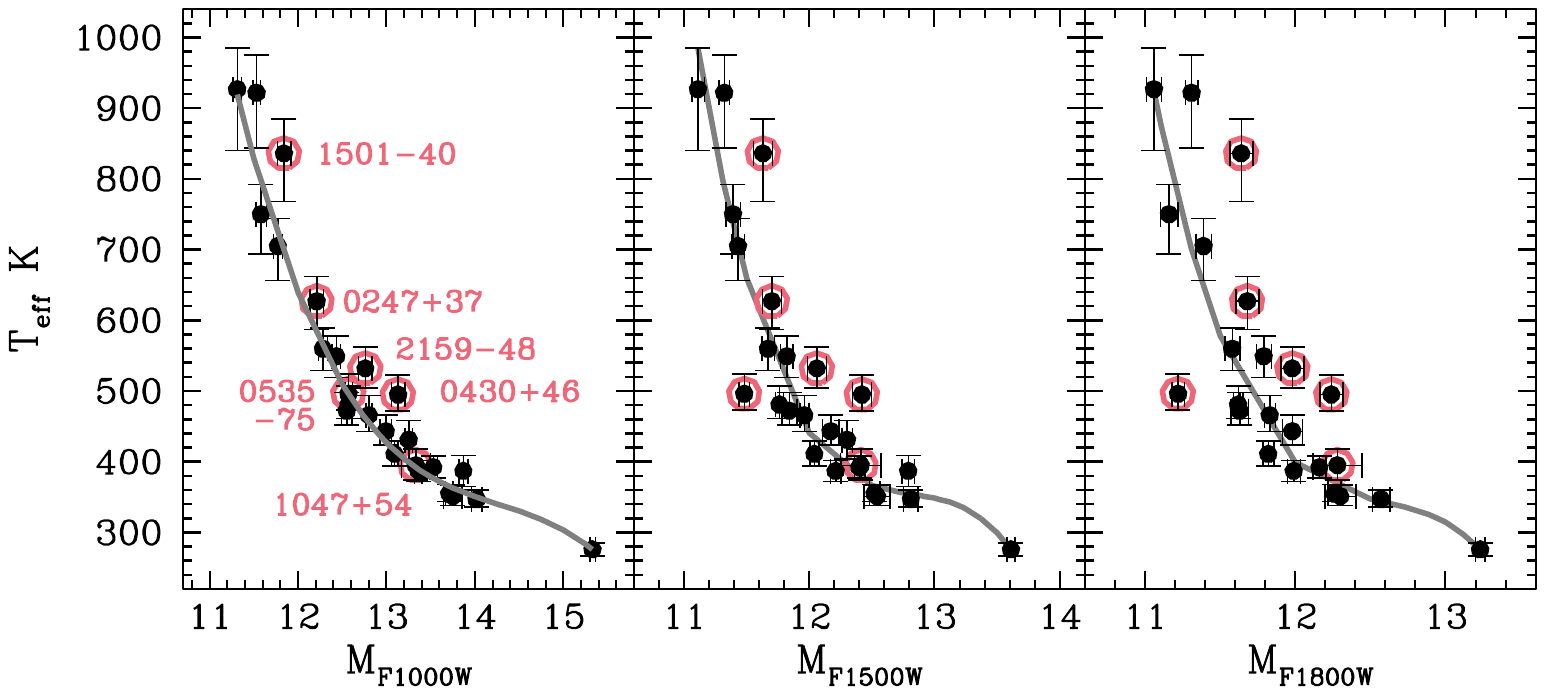}
\vskip -0.2in
\caption{Luminosity-based $T_{\rm eff}$ for the sample of 25 T and Y dwarfs described in Section 2, as a function of absolute magnitudes at $\lambda \approx 10, 15$ and $18~\mu$m.  Six brown dwarfs are omitted from the regression fits, identified left panel, and these points are circled in red.   The results of the fits are shown as gray lines and the parameters are given in Table 3.}
\end{figure}

\begin{deluxetable*}{crrrrcc}
\tablecaption{Coefficients of Fits to Color:$T_{\rm eff}$ Relationships}
\tablehead{\\[0.01in]
\colhead{Color, $x$} &  \colhead{a0} &  \colhead{a1} &  \colhead{a2} &  \colhead{a3}  & \colhead{$rms$~(K)} & \colhead{Valid for}\\
}
\startdata
$J -$ W2 & 1516.78 & -383.075 & 43.7989 & -1.73969 & 16   & $1.99 \leq J -$ W2 $\leq 11.20$   \\
$J -$ [4.5] & 1507.82 & -381.645 &  43.9661 &  -1.76383 & 16  & $2.01 \leq J -$ [4.5] $\leq 11.10$ \\
{[}3.6] $-$ [4.5] & 2302.47 & -1735.962 &  550.974 & -62.7196 & 13 & $1.00 \leq$ [3.6] $-$ [4.5] $\leq 3.55$\\
$M_{W2}$ & 56119.60 &  -10298.389 & 637.2170 & -13.21476 & 13 & $12.92 \leq M_{W2} \leq 17.03$\\
$M_{[4.5]}$ &  54076.96 &  -9857.745 & 605.8006 & -12.47410 & 13 &  $12.89 \leq M_{[4.5]} \leq 17.13$ \\
$M_{F480M}$ &  74390.62 & -13814.799 &  862.5367 & -18.02330 & 13 & $13.04 \leq M_{F480M} \leq 16.89$  \\
$M_{F444W}$ &  42005.61 & -7520.306 & 455.4078 &  -9.25657 & 12 &  $12.73 \leq M_{F444W} \leq 17.29$ \\
$M_{F1000W}$ & 52562.93 &  -10898.188 & 760.8665 &  -17.77239 & 10 & $11.31 \leq M_{F1000W} \leq 15.33$ \\
$M_{F1500W}$ & 246853.60 &  -57618.770 & 4491.4848 & -116.76018 & 14 & $11.11 \leq M_{F1500W} \leq 13.61$ \\
$M_{F1800W}$ & 286936.96 & -68205.344 &  5414.2448 &  -143.35935 & 19 & $11.06 \leq M_{F1800W} \leq 13.23$ \\
\enddata
\smallskip
\tablecomments{
Coefficients are applied as  
$$
T_{\rm eff} = a0 + a1 \times x + a2 \times x^2 + a3 \times x^3  
$$}
\tablenotetext{*}{The $rms$ of each fit is given in the Table. The  total uncertainty in $T_{\rm eff}$ determined from these relationships is estimated to range from 20~K to 60~K for the colder to warmer brown dwarfs; see Section 3.2. }
\end{deluxetable*}

For late-T and Y dwarfs, absolute magnitudes at $\lambda \sim 4.5~\mu$m can provide a good estimate of $T_{\rm eff}$
\citep[e.g.][]{Leggett_2025a, Leggett_2025b}, while  colors such as $J -$ [4.5] and [3.6] $-$ [4.5] are more sensitive to metallicity and gravity \citep[e.g.][]{Meisner_2020a,Leggett_2025a, Leggett_2025b}.  Table 3 gives the dispersion in the fits shown in Figures 2, 3, and 4. The largest dispersion is for $M_{F1800W}$, followed by the $J$-band colors. All dispersions are low, however this is for a small sample of typical field age and metallicity brown dwarfs. 
Applying the $J-$ W2, $J-$ [4.5], [3.6] $-$ [4.5], $M_{\rm W2}$, and $M_{\rm [4.5]}$ relationships, or a subset, to 
a photometric data set of 308 dwarfs cooler than 1000~K, produced standard deviations ($s$) in $T_{\rm eff}$ which increased with $T_{\rm eff}$.  This reflects the increasing steepness of the color:$T_{\rm eff}$ curve with increasing $T_{\rm eff}$ (Figures 2, 3, and 4). After $3\sigma$-clipping, for $T_{\rm eff} \approx 350$~K, $s = 20$~K; for $T_{\rm eff} \approx 550$~K, $s = 41$~K; and for $T_{\rm eff} \approx 750$~K, $s = 57$~K. These values, of $\approx 7\%$ in $T_{\rm eff}$, represent the uncertainty floor when using the relationships in Table 3.   Extreme outliers in metallicity or gravity can have an even larger dispersion in $T_{\rm eff}$ across the colors, as shown in the Section 3.3.

\subsection{Atypical Brown Dwarfs}

The six sources excluded from the regression fits, and circled in red in Figures 2, 3, and 4, are:
\begin{itemize}
    \item WISE J024714.52$+$372523.5: red in [3.6] $-$ [4.5] and the spectrum shows weak CO$_2$ and CO for a T8 dwarf \citep[Figure 4 of][]{Tu_2024}. This object may be metal-poor or have a high gravity \citep{Tu_2024, Leggett_2025a}. 
    \item WISE J043052.92$+$463331.6: appears faint at long wavelengths (Figure 4) and bright at $\lambda \approx 4.5~\mu$m compared to other T8 dwarfs \citep{Beiler_2024, Tu_2024}. It has the bluest F444W $-$ F1000W color in the  \citet{Beiler_2024} sample, and has been flagged as abnormally red in $H -$ W2 and faint in $H$ by \citet{Best_2021}.  \citet{Beiler_2024} and \citet{Tu_2024} also flag this brown dwarf as peculiar in their analyses of the {\it JWST} data.  It is currently unclear what could produce such an SED.
    \item WISE J053516.80$-$750024.9 (hereafter WISE 0535-75): has extremely weak 
    CO$_2$ and CO absorption \citep{Beiler_2024, Tu_2024}, leading to very red 
    $J -$ W2, $J -$ [4.5], and [3.6] $-$ [4.5] colors.
    This object may be very metal-poor and/or have a high gravity, or it may be a binary system.  In a preliminary comparison of the {\it JWST} spectra to ATMO 2020++ synthetic spectra, \citet{Leggett_2024}
    found that reasonable fits could be obtained with a $T_{\rm eff} = 500$~K, log $g = 4.5$, [m/H] $= -0.5$ synthetic spectrum.  A slightly better fit was obtained adopting a binary solution of
    450~K and 350~K solar-metallicity brown dwarfs. Using the color relationships determined here, I find that the $J -$ [4.5] and [3.6] $-$ [4.5] colors, and the absolute [4.5], F1000W, F1500W, and F1800W magnitudes, are well reproduced if the system is a binary composed of a 480~K and 340~K Y dwarf pair.  This is illustrated in Figure 2 for the $J -$ [4.5] color.  
    \item CWISEP J104756.81$+$545741.6 (hereafter WISE 1047+54): blue $J -$ W2 and $J -$ [4.5], with strong  CO$_2$ and CO \citep[Figure 4 of][]{Tu_2024}. This object is very likely to be a young, metal-rich, low-gravity and low-mass  dwarf \citep{Beiler_2024, Leggett_2025b}.
    \item WISE J150115.92$-$400418.4: sub-luminous at 
    $\lambda \approx 4.5~\mu$m, with strong CO$_2$ and CO \citep[Figure 4 of][]{Tu_2024}. This T dwarf is likely to be metal-rich or have a low gravity.
    \item WISEA J215949.54-480855.2: red in [3.6] $-$ [4.5] and the spectrum shows weak CO$_2$ and CO for a T9 dwarf \citep[Figure 4 of][]{Tu_2024}. This object may be metal-poor or have a high gravity; the relatively high tangential velocity of 82~km~s$^{-1}$ \citep{Kirkpatrick_2021a} may support membership of an older disk population  \citep[their Figure 31]{Dupuy_2012}.
\end{itemize}

\bigskip
\section{The Coolest Y Dwarfs}

Table 4 includes all cold solar neighborhood Y dwarfs from the literature that the author is aware of (as of February 2026), with $T_{\rm eff} < 425$~K. 
Luminosity-based $T_{\rm eff}$ values are given, where they exist. $T_{\rm eff}$ values are also given for available colors using the relationships of Table 3.  The uncertainties in the color values are the square root of the sum of the squares of the $rms$ for the relationship (Table 3) and the $\delta T_{\rm eff}$ due to the uncertainty in the color.  The photometry is taken from Table 2 and \citet{Leggett_2025_erratum}.

The adopted $T_{\rm eff}$ values in Table 4 are equal to the luminosity-based values  where available.  Otherwise the  adopted $T_{\rm eff}$ values are the weighted mean of the color values, where I have first 
averaged the results from $J -$ W2 and $J -$ [4.5], and the results from $M_{W2}$, $M_{[4.5]}$, and $M_{F480M}$, as these sample the same region of the SED.  The quoted uncertainty in the color-based $T_{\rm eff}$ values is the uncertainty in the weighted mean or 20~K, whichever is larger (see Section 3.2).  The color-based $T_{\rm eff}$ values are within $2\sigma$ of the adopted values except for WISE 1047+54, previously discussed and thought to be unusually low-gravity and metal-rich.  For CWISEP J225628.97+400227.3 the $T_{\rm eff}$ values determined from $J -$ W2 and $J -$ [4.5] are 70~K higher than the values determined from [3.6] $-$ [4.5], $M_{W2}$, and $M_{[4.5]}$ and I exclude them from the mean; this object is faint at $J$ and it is possible that the uncertainty in the $J$ measurement by \citet{Meisner_2020a}, although already large, has been underestimated.

Three of the Y dwarfs do not have a measured parallax:
WISE J021243.55+053147.2, WISE J125721.01+715349.3, and WISE J235120.62-700025.8.  A photometric distance is given in Table 4 in square brackets based on the 
 the observed  $\lambda \sim 5~\mu$m magnitudes and the  absolute magnitude indicated by the $T_{\rm eff}$ value.  The uncertainty in the photometric distance due to the scatter in the $T_{\rm eff}$-color relationships is $\approx 10\%$.  A photometric distance is also given for  WISE J223022.60+254907.5 which has a very uncertain parallax.

The distant candidate Y dwarfs identified in the JADES survey  \citep{Hainline_2024} are not included in Table 4.
\citet[][their Table 6]{Hainline_2025} 
lists three candidates cooler than 400~K, with $325 \lesssim T_{\rm eff}$~K $\lesssim 380$:
JADES-GS-BD-5 at a distance of $\sim 80$~pc,
JADES-GS-BD-21 at a distance of $\sim 800$~pc,
and JADES-GN-BD-16 at a distance of $\sim 900$~pc.
\citet{Hainline_2025} uses various {\it JWST} color criteria to select candidate Y dwarfs. In the Appendix I explore these colors further.

\startlongtable
\begin{deluxetable*}{lcccccccccccc}
\setlength{\tabcolsep}{2pt}
\tabletypesize{\scriptsize}
\tablecaption{Candidate Brown Dwarfs with $T_{\rm eff} < 425$~K}
\tablehead{
\\[0.0001in] 
\colhead{WISE J or} & \colhead{Distance} & 
 \multicolumn{10}{c}{$T_{\rm eff}$~(K)}  & \colhead{Ref.}\\
 \colhead{Other Name}  & \colhead{pc} &
 \colhead{Lumin.} &
 \colhead{$J -$ W2} &   \colhead{$J -$ [4.5]} &  \colhead{[3.6] $-$ [4.5]} & 
 \colhead{$M_{W2}$} & \colhead{$M_{[4.5]}$} & \colhead{$M_{F480M}$} & \colhead{$M_{F1000W}$} & \colhead{$M_{F1500W}$} & \colhead{Adopted} & \colhead{} 
}
\startdata
014656.66 &  19.34$\pm$0.72 &  & & 429$\pm$16 &  408$\pm$16 &  & 436$\pm$25 &  & 373$\pm$20 & & 411$\pm$20 & 1,2,18,\\
+423410.0B & &&&&&&&&&&&26\\
021243.55  & [13.0$\pm$1.3] &  &  394$\pm$16    & 394$\pm$16 &  372$\pm$32 &  &  &  &  &  &391$\pm$20 & 3,26\\
+053147.2& &&&&&&&&&&&\\
033605.05  & 10.02$\pm$0.22 &   &      &      &  &   &   &  308$\pm$26  &  & & 308$\pm$26  & 4,5,18,\\ 
-014350.4B & &&&&&&&&&&&22\\
035000.32  & 5.67$\pm$0.07   &  & 376$\pm$16  & 377$\pm$16 & 327$\pm$17 & 349$\pm$13 &    349$\pm$13 & &   &   &   355$\pm$20 & 1,18,27\\
-565830.2 & &&&&&&&&&&&\\
040235.55 & 8.6$^{+1.8}_{-1.3}$  &  & 358$\pm$18 & 356$\pm$18 & 395$\pm$22 &  350$\pm$29 & 356$\pm$28 &  & & & 364$\pm$20 & 3,19,26\\
-265145.4 & &&&&&&&&&&&\\
041022.71 & 6.54$\pm$0.03 &   & 455$\pm$16    & 460$\pm$16    & 431$\pm$13 & 416$\pm$13 & 407$\pm$13 & 420$\pm$14 & 405$\pm$11 & 388$\pm$13 & 420$\pm$20 & 6,20,28\\
+150248.4& &&&&&&&&&&&29\\
050305.68& 10.2$\pm$0.4 &  & 398$\pm$16 & 400$\pm$16 & 362$\pm$25 &  348$\pm$15 & 345$\pm$14 & &  &  &  364$\pm$20 & 7,19,26\\
-564834.0 & &&&&&&&&&&&\\
053516.80 & 14.56$\pm$0.43 &   &  & & & & & & & & 340$\pm$20* & 1,19,30\\
-750024.9B & &&&&&&&&&&&\\
064723.23& 10.05$\pm$0.17 &  & 367$\pm$16 & 368$\pm$16 & 382$\pm$16 & 408$\pm$14 & 410$\pm$14 & 421$\pm$15 &  & & 394$\pm$20 & 8,19,18,\\
-623235.5 & &&&&&&&&&&&26,27,31\\
WD 0806-661B & 19.23$\pm$0.01 &  & 353$\pm$16 & 355$\pm$16 & 426$\pm$15 & 392$\pm$17 & 383$\pm$16 &   &  &  & 383$\pm$20 & 9,21,27,31\\
082507.35 & 6.55$\pm$0.09  & 387$\pm$15 & 365$\pm$16 & 365$\pm$16 & 400$\pm$14  &  370$\pm$13 & 369$\pm$13 & 373$\pm$14 & 390$\pm$11& 397$\pm$17 & 387$\pm$15 & 10,19,18,\\
+280548.5& &&&&&&&&&&&30\\
083011.95 & 10.08$^{+0.71}_{-0.62}$  &  & & &326$\pm$22 & 346$\pm$18 & 354$\pm$16  & 362$\pm$17 & & &  345$\pm$20 & 11,22,26\\
+283716.0& &&&&&&&&&&&\\
085510.83 & 2.28$\pm$0.01 & 276$\pm$9 & 276$\pm$15  & 276$\pm$16  & 277$\pm$16  & 276$\pm$14 & 276$\pm$13  &  276$\pm$14 & 276$\pm$10 & 276$\pm$17 & 276$\pm$9 & 12,18,30\\
-071442.5& &&&&&&&&&&&\\
094005.50& 15.06$^{+3.39}_{-2.33}$  &  & 421$\pm$16 & 414$\pm$16 & 388$\pm$23 &  416$\pm$49 & 431$\pm$53 &  &   &   & 413$\pm$20 & 3,19,26\\
+523359.2 & &&&&&&&&&&&\\
MEAD 62B** & 20.48$\pm$0.01  &  &      &      &  &   & & & 294$\pm$11 & 312$\pm$15 & 299$\pm$20 & 13,23\\  
104756.81 & 14.7$^{+1.1}_{-1.0}$ & 395$^{+23}_{-21}$ & 476$\pm$16  & 481$\pm$16  & 431$\pm$28  & 387$\pm$21  & 378$\pm$19  & 368$\pm$18  & 393$\pm$20  & 374$\pm$23  & 395$^{+23}_{-21}$ & 3,24,30\\
+545741.6& &&&&&&&&&&&\\
125721.01 & [14.0$\pm$1.4] &  &  378$\pm$16  & 381$\pm$16  & 394$\pm$23  &   &  &    &   & &  382$\pm$20 & 7,26\\
+715349.3& &&&&&&&&&&&\\
140518.40 & 6.32$\pm$0.10  & 392$^{+16}_{-15}$ & 386$\pm$16  & 386$\pm$16  & 382$\pm$14  & 408$\pm$14  & 410$\pm$14  & 413$\pm$15  & 395$\pm$11  & 374$\pm$15  & 392$^{+16}_{-15}$ & 6,19,3,\\
+553421.4& &&&&&&&&&&&28,29\\
144606.62 & 9.63$^{+0.49}_{-0.44}$  & 351$^{+16}_{-13}$ & 386$\pm$16  & 386$\pm$16  & 357$\pm$13  & 343$\pm$16  & 344$\pm$14  & 346$\pm$15  
&  363$\pm$12  & 364$\pm$16  & 351$^{+16}_{-13}$ & 3,24,26,\\
-231717.8& &&&&&&&&&&&30\\
154151.66 & 6.02$\pm$0.04 &  411$^{+18}_{-17}$ & 384$\pm$15  & 383$\pm$16  & 385$\pm$13  & 388$\pm$14  & 394$\pm$13  & 393$\pm$13  & 415$\pm$11  & 429$\pm$19  & 411$^{+18}_{-17}$ & 6,18,28,\\
-225025.2& &&&&&&&&&&&30\\
163940.86 & 4.74$\pm$0.01  &  & 389$\pm$16  & 389$\pm$16  & 411$\pm$13  & 395$\pm$13  & 396$\pm$13  & 400$\pm$14  &  &   &  398$\pm$20 & 14,25,18,\\
-684744.6& &&&&&&&&&&&31\\
182831.08 & 9.97$\pm$0.19 & 387$\pm$22 & 354$\pm$16  & 353$\pm$16  & 411$\pm$13  & 403$\pm$14  & 408$\pm$14  & 411$\pm$15  & 357$\pm$11  & 355$\pm$14  &387$\pm$22*** & 6,19,18,\\
+265037.7A,B& &&&&&&&&&&&30\\
193054.55 & 9.4$\pm$0.4 &  & 380$\pm$16 & 
379$\pm$16 & 338$\pm$17 & 377$\pm$16 &  382$\pm$16 &
& & & 372$\pm$20 & 7,19,26\\
-205949.4& &&&&&&&&&&&\\
193518.58 & 14.4$\pm$0.8 &  &      &   &  & & & & 404$\pm$18 & & 404$\pm$20 & 15,16,19\\
-154620.3A & &&&&&&&&&&&\\
193518.58 & 14.4$\pm$0.8 &  &      &   &  &  & & & 325$^{+16}_{-34}$ & &325$^{+20}_{-34}$ & 15,16,19\\ 
-154620.3B  & &&&&&&&&&&&\\
220905.73& 6.18$\pm$0.08 & 355$^{+13}_{-11}$ & 361$\pm$16 & 362$\pm$16 & 350$\pm$15 & 358$\pm$13 & 356$\pm$13 & 356$\pm$14 & 365$\pm$11 & 365$\pm$14 & 355$^{+13}_{-11}$ & 17,19,18,\\
+271143.9 & &&&&&&&&&&&31\\
223022.60 & 14$^{+4}_{-3}$ [15.0$\pm$1.5]  &  & 391$\pm$16 & 394$\pm$17 & 381$\pm$36 & 376$\pm$45 & 370$\pm$39 & &  &   & 389$\pm$20 & 3,19,26\\
+254907.5& &&&&&&&&&&&\\
225628.97 &  9.8$^{+1.2}_{-1.0}$  &  & (427$\pm$24) & (425$\pm$25) & 359$\pm$30 & 351$\pm$21 & 352$\pm$19 &  &   &  &  353$\pm$20 & 3,19\\
+400227.3& &&&&&&&&&&&\\
235120.62 & [9.0$\pm$0.9] &  &  390$\pm$17 & 393$\pm$17 &396$\pm$18 &  &  &  &  &  &  393$\pm$20 & 7,32\\
-700025.8& &&&&&&&&&&&\\
235402.79 & 7.66$\pm$0.20  & 347$^{+13}_{-11}$ & 366$\pm$16 & 367$\pm$16 & 347$\pm$17 & 369$\pm$14 & 368$\pm$14 & 363$\pm$14 & 350$\pm$11 & 354$\pm$14 &  347 $^{+13}_{-11}$ & 10,19,18,\\
+024014.1& &&&&&&&&&&&30\\
\enddata
\tablenotetext{*}{~Assuming that this super-luminous object is a binary with component $T_{\rm eff}$ values of 480~K and 340~K, as deconvolved with the photometry in Section 3.}
\tablenotetext{**}{~~Unconfirmed companion to 2MASS J09424023-46371 \citep{Albert_MEAD_2025}.}
\tablenotetext{***}{~~~Assuming that this super-luminous object is composed of an identical pair of Y dwarfs \citep[e.g][]{Leggett_2025a}. }
\tablerefs{DISCOVERY: 
1 -- \citet{Kirkpatrick_2012};
2 -- \citet{Dupuy_2015};
3 -- \citet{Meisner_2020a};
4 -- \citet{Mace_2013a};
5 -- \citet{Calissendorff_2023}
6 -- \citet{Cushing_2011};
7 -- \citet{Meisner_2020b};
8 -- \citet{Kirkpatrick_2013};
9 -- \citet[][aka WISE J080714.68-661848.7]{Luhman_2011};
10 -- \citet{Schneider_2015};
11 -- \citet{Bardalez_2020};
12 -- \citet{Luhman_2014};
13 -- \citet[][aka 2MASS J09424023-4637176B]{Albert_MEAD_2025};
14 -- \citet{Tinney_2012};
15 -- \citet{Marocco_2019};
16 -- \citet{Furio_2025};
17 -- \citet{Cushing_2014}.
PARALLAX: 
18 --   \citet{Kirkpatrick_2019};
19 -- \citet{Kirkpatrick_2021a};
20 -- \citet{Fontanive_2025}
21 -- \citet{Subasavage_2009};
22 -- \citet{Albert_2025};
23 -- \citet{GAIA};
24 -- \cite{Beiler_2024};
25 -- \citet{Fontanive_2021}.
PHOTOMETRY:
2 -- \citet{Dupuy_2015};
3 -- \citet{Meisner_2020a};
7 -- \citet{Meisner_2020b};
13 -- \citet[][aka 2MASS J09424023-4637176B]{Albert_MEAD_2025};
18 --   \citet{Kirkpatrick_2019};
22 -- \citet{Albert_2025};
26 -- \citet{Leggett_2021};
27 -- \citet{Leggett_2015};
28 -- \citet{Leggett_2013};
29 -- \citet{Leggett_2019};
30 -- \citet{Leggett_2025b};
31 -- \citet{Leggett_2017};
32  -- \citet{Kirkpatrick_2024}.
}
\end{deluxetable*}

\begin{figure}
\vskip -0.2in
\hskip 0.5in
\includegraphics[angle=0, width = 5.0 in]{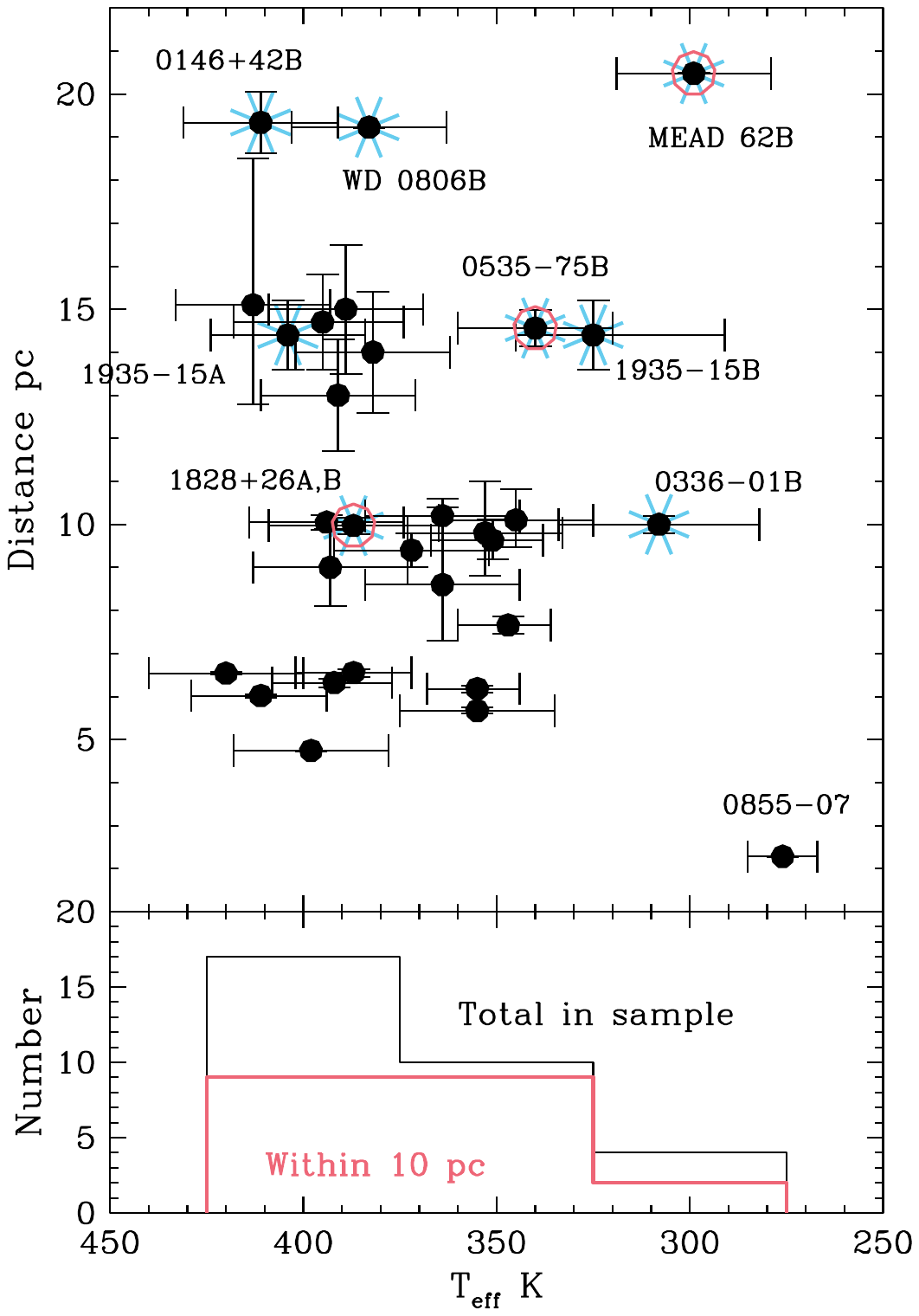}
\vskip -0.5in
\caption{The cold solar neighborhood Y dwarfs from Table 4. Data points with red circles are unconfirmed. Data points with blue asterisks are components of binary systems.
}
\end{figure}

Figure 5 shows the cold Y dwarfs listed in Table 4 in a $T_{\rm eff}$:distance plot and in a   histogram as a function of $T_{\rm eff}$.   There are 31 known objects with $T_{\rm eff} < 425$~K if WISE 0535-75 and
WISE 1828+26 are unresolved binary systems, and if MEAD 62B is proven to be a Y dwarf companion to the white dwarf MEAD 62.  Note that WISE 0336-01A is not included in Table 4 or Figure 5; \citet{Calissendorff_2023} estimate $T_{\rm eff} = 415 \pm 20$~K for WISE 0336-01A however using the relationships presented here I find a warmer value for this brown dwarf of  $T_{\rm eff} = 446 \pm 20$~K.

The total number with $275 < T_{\rm eff}$~K $< 425$ within 10~pc (which is more likely to be complete than the $\sim 20$~pc sample)  is 20, or 18 if WISE 1828+26 is a single, warmer, brown dwarf.  This translates into a number density of between $4\times 10^{-3}$pc$^{-3}$ and $6\times 10^{-3}$pc$^{-3}$ for Y dwarfs across the 150~K range $275 < T_{\rm eff}$~K $< 425$, consistent with the accounting of nearby objects by \citet[][their Table 17]{Kirkpatrick_2024}.  
The coldest bin, with the least luminous objects, may be incomplete.  Simulations by \citet{Kirkpatrick_2024}  assuming a constant star formation rate with low-mass cut-offs of 1, 5, or 10 $M_{\rm Jupiter}$, imply that the number of objects with $150 < T_{\rm eff}$~K $< 300$ ranges from $\approx 0.2 \times 10^{-3}$pc$^{-3}$ using \citet{Saumon_2008} evolutionary models, to $\approx 8  \times 10^{-3}$pc$^{-3}$ using \citet{Baraffe_2003} models combined with a low-mass cut-off of 1 $M_{\rm Jupiter}$.  If the former number density is valid, then we may not find any other Y dwarfs like WISE 0855-07 within 10~pc of the Sun.

\bigskip
\section{The Cold Binary Systems}

Figure 6 shows evolutionary sequences for T dwarfs in the form of a $T_{\rm eff}$:log~$g$ diagram. The evolutionary model is from \citet{Marley_2021}. Blue regions indicate the locations of the WISE 0336-01 components, and red regions the WISE 1935-15 components. The green box is MEAD 62B where the age has been constrained to be equal to that of the white dwarf, 4.5 -- 10.8~Gyr \citep{Albert_MEAD_2025}.  This constraint on age, combined with our $T_{\rm eff}$ estimate, indicates a range in mass for MEAD 62B of 9 to 12 $M_{\rm Jupiter}$, if it is indeed a companion to the white dwarf.

\begin{figure}
\hskip 0.5in
\vskip -0.4in
\includegraphics[angle=-90, width = 6.8 in]
{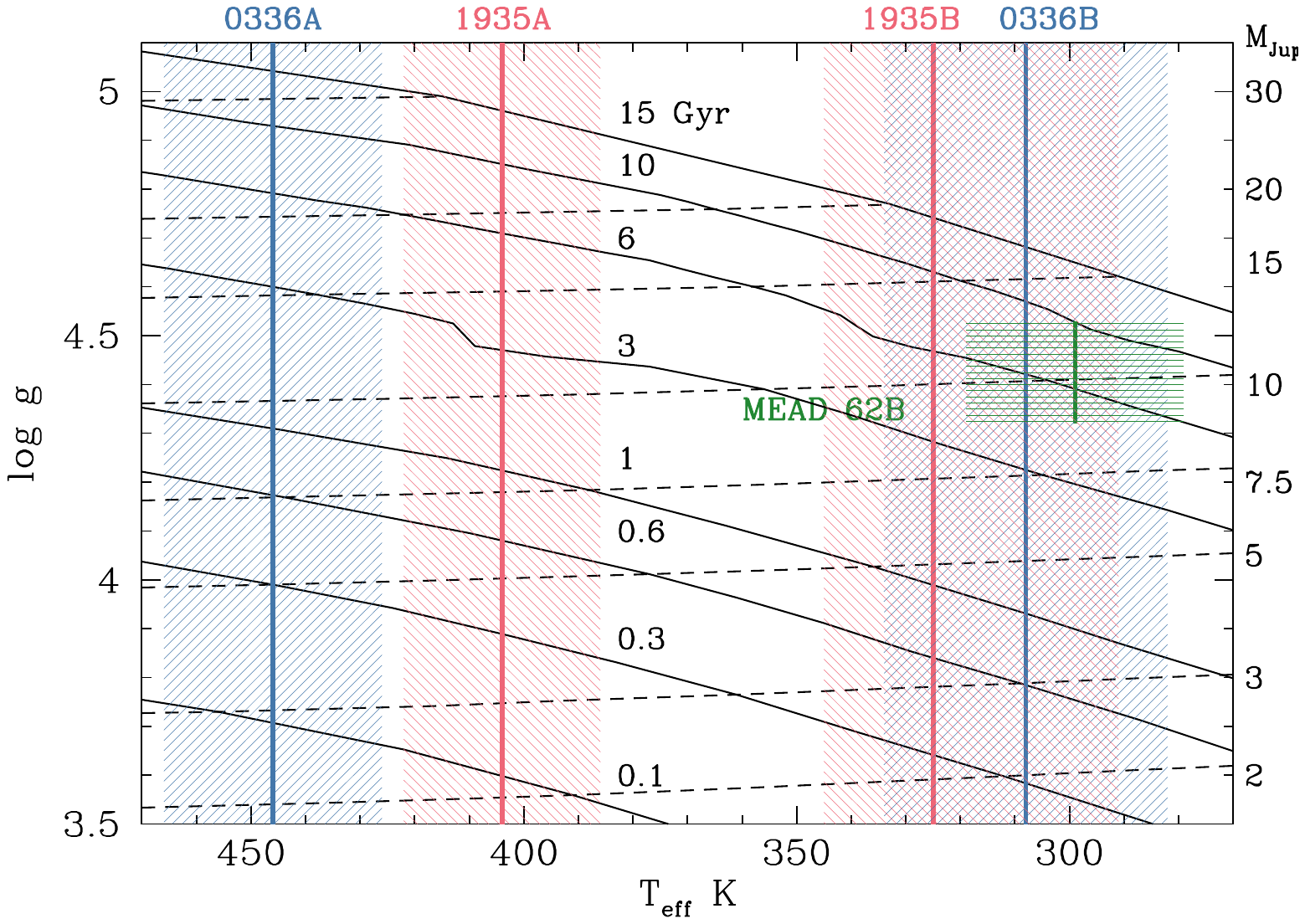}
\vskip -0.2in
\caption{Evolutionary diagram for Y dwarfs from models by \citet{Marley_2021}. Solid black lines are constant age, with age in Gyr above each line. Dashed lines are iso-mass sequences with mass given on the right axis.  Blue vertical line and hashed regions indicate the $T_{\rm eff}$ values and their uncertainties for the WISE 0336-01 binary system. Red lines and regions indicate the same for the WISE 1935-15 binary system. The green line and rectangular region illustrate the   $T_{\rm eff}$ and uncertainty for MEAD 62B, restricting the age to be that of the white dwarf primary, 4.5 -- 10.8~Gyr \citep{Albert_MEAD_2025}.
}
\end{figure}

\begin{deluxetable*}{cccccccccl}[b]
\setlength{\tabcolsep}{2pt}
\tabletypesize{\footnotesize}
\tablecaption{Binary Systems with Cold Y Dwarf Secondaries}
\tablehead{
\colhead{System} &   &
\multicolumn{2}{c}{Primary} & \multicolumn{2}{c}{Secondary} &
     \multicolumn{2}{c}{Separation} & \\
\colhead{Name}  & \colhead{Age (Gyr)}  &
\colhead{$T_{\rm eff}$~(K)}  & \colhead{Mass ($M_{\rm Jup}$)}  & 
 \colhead{$T_{\rm eff}$~(K)}  & \colhead{Mass ($M_{\rm Jup}$)}  &
  \colhead{mas} & \colhead{au} &
\colhead{Reference} 
}
\startdata
WISE 0336-01 & 0.5 -- 10.0 & 446 $\pm$ 20 & 6 -- 28  & 308  $\pm$ 26 & 2 -- 17  & 84 & 0.97 & \citet{Calissendorff_2023}\\
MEAD 62* & 6.64 $^{+4.15}_{-2.14}$ & 5968  $\pm$ 124 & 628 $\pm$ 20  & 299 $\pm$ 20 & 9 -- 12  & 1953 & 40.0 & \citet{Albert_MEAD_2025} \\
WISE 1935-15 & 0.5 -- 10.0 & 404 $\pm$ 20 &  5 -- 25 &  325 $^{+20}_{-34}$ & 2 -- 18  & 172 & 2.48 & \citet{Furio_2025}\\
\enddata
\tablenotetext{*}{Also known as 2MASS J09424023-4637176. The primary is a magnetic DA white dwarf \citep{OBrien_MEAD_WD}; the candidate Y dwarf secondary has not been confirmed to be associated with the white dwarf  however the colors and luminosity are consistent with a 299~K dwarf at the distance of the white dwarf.}
\tablecomments{The ages of the Y dwarf binary systems are assumed to be between 0.5 and 10~Gyr, typical of a local population \citep[e.g.][]{Buder_2019}.  
The age of the white dwarf system and the white dwarf temperature is from \cite{Albert_MEAD_2025}.  The brown dwarf $T_{\rm eff}$ values are photometric estimates using the relationships presented in this work.  Masses for the brown dwarfs are determined using age and $T_{\rm eff}$ and the \citet{Marley_2021} evolutionary models. The mass of the white dwarf is from \cite{Albert_MEAD_2025}.}
\end{deluxetable*}

The ages of the WISE 0336-01 and WISE 1935-15 systems are not constrained, but I assume that each binary is coeval. Adopting an  age range of 0.5 to 10 Gyr for a local dwarf population
\citep[e.g.][]{Buder_2019} then the range in masses are: 6 -- 28 and 2 -- 17  $M_{\rm Jupiter}$ for WISE 0336-01 A and B, and 5 -- 25 and 2 -- 18  $M_{\rm Jupiter}$ for WISE 1935-15 A and B. The range in mass correlates with the range in age, that is, the younger the system the smaller the mass and vice versa (Figure 6). 
Note that the population synthesis of brown dwarfs within 25~pc by \citet{Best_2024} shows that more than 90\% of the population is younger that 4~Gyr, implying that the masses of the cold binary components are unlikely to be at the high end of their ranges. 
For a solar-like age, the component masses are 19 and 9 $M_{\rm Jupiter}$ for WISE 0336-01 A and B, and 17 and 11 $M_{\rm Jupiter}$ for WISE 1935-15 A and B (Figure 6).
The properties of the three systems are summarized in Table 5.

\bigskip
\section{Conclusions}

A decade after the discovery of WISE 0855-07 \citep{Luhman_2014}, three objects have been found which may rival its extremely low temperature and luminosity. Two of these are secondary components of Y dwarf$+$Y dwarf binary systems: WISE 0336-01B and WISE 1935-15B are in tight configurations with warmer Y dwarfs \citep{Calissendorff_2023, Furio_2025}.  The third object, MEAD 62B, needs to be confirmed as a true companion, but it appears to be a cold Y dwarf in  a wider configuration with a white dwarf primary \citep{Albert_MEAD_2025}. 

In this work I have determined relationships between mid-infrared colors and $T_{\rm eff}$, where the $T_{\rm eff}$ values have been constrained by measurements of bolometric luminosity.  For very cold 
Y dwarfs a measurement of luminosity provides tight constraints on $T_{\rm eff}$ (Figure 1).  Luminosity-based  $T_{\rm eff}$ values have been taken from \cite{Beiler_2024} and determined here (Section 2). 

Weighted regression fits showed that cubic polynomials reproduced color trends with $T_{\rm eff}$ well, once $3\sigma$ outliers were removed.  The $rms$ of the fits is on average 14~K, however a comparison of $T_{\rm eff}$ values for each color for a larger sample of T and Y dwarfs suggests that the uncertainty floor is $\approx 60$~K for 750~K late-T dwarfs, $\approx 40$~K for 550~K late-T/early Y dwarfs, and $\approx 20$~K for 350~K Y dwarfs.

Six atypical brown dwarfs were excluded from the regression fitting. Two have the spectral signatures of metal-paucity and/or increased gravity (and hence age): WISE J024714.52+372523.5 and WISEA J215949.54-480855.2.  Two have the spectral signatures of metal enrichment and/or decreased gravity (and hence age): 
 CWISEP J104756.81+545741.6 and WISE J150115.92-400418.4.  WISE J053516.80-750024.9 may be metal-poor and high-gravity, but is better fit as an unresolved binary consisting of 480~K and 340~K Y dwarfs. WISE J043052.92+463331.6 has a peculiar SED which cannot currently be interpreted.

The color:$T_{\rm eff}$ relationships  determined here were used to estimate the $T_{\rm eff}$ values
for the coldest known Y dwarfs, including the candidate cold Y dwarf secondaries in known or suspected binary systems.  I built a sample of 31 objects with $275 \leq T_{\rm eff}$~K $\leq 425$  at distances of about 2 to 20~pc (Figure 5); this assumes that
WISE 0535-75 and WISE 1828+26 are unresolved binary systems (Sections 2 and 3), and MEAD 62B is  a Y dwarf companion to the white dwarf \citep{Albert_MEAD_2025}.   I find $299 \leq T_{\rm eff} \leq 325$ for WISE 0336-01B, WISE 1935-15B, and the candidate MEAD 62B, and evolutionary models  indicate that their masses are $\sim 10~M_{\rm Jupiter}$ if the systems are a few Gyrs old \citep{Marley_2021}.
The  sample provides a high priority target list for  {\it JWST} observations, given that it covers the critical regime where water clouds form in the photosphere \citep{Morley_2014}.

Cool brown dwarfs are now being found at kpc distances in {\it JWST}  data \citep[e.g.][]{Burgasser_2024_UNCOVER, Hainline_2024, Hainline_2025, Tu_2025_distant}. In the Appendix I give measured and synthesized {\it JWST} colors for a sample of brown dwarfs, including the extremely metal-poor 500~K object  WISEA J153429.75-104303.3, also known as ``The Accident'' \citep{Kirkpatrick_2021b, Faherty_2025}. These colors will be useful references for deep searches for cold brown dwarfs in {\it JWST} data, which may probe metal-poor populations in the Milky Way.

\bigskip
\begin{acknowledgements}
The author is grateful to the anonymous referee whose comments greatly improved this manuscript.

\end{acknowledgements}

\bigskip
\appendix

\section{{\it JWST} Colors for Identifying Distant Cold Brown Dwarfs}

Recently, distant candidate brown dwarfs have been found in deep searches of {\it JWST} data \citep[e.g.][]{Burgasser_2024_UNCOVER, Hainline_2024, Hainline_2025, Tu_2025_distant}.  In order to classify these objects, synthetic colors calculated by model atmospheres are generally used.
In this section I provide {\it JWST} observational colors for a sample of brown dwarfs which can be used as a reference for the distant population.   I use the five filters of the JADES survey \citep{Hainline_2025}, which overlap with the filters used by the UNCOVER survey \citep{Burgasser_2024_UNCOVER}: F115W, F150W, F277W, F410M, and F444W.

The primary data source is the \citet{Beiler_2024} sample of T and Y dwarfs with luminosity-based $T_{\rm eff}$ values.  \citet{Beiler_2024} synthesize 
F115W, F150W, F277W, F410M, and F444W colors using their low-resolution spectra.  I add three objects to this sample: WISE 0855-07, WISEA J153429.75-104303.3 (hereafter WISE 1534-10), and WISE 1828+26. The photometry is synthesized from the NIRSpec spectra published by \citet{Luhman_2024} and \citet{Faherty_2025}, and the spectra obtained for WISE 1828+26 as part of the JWST Cycle 1 GTO program 1189, PI: Thomas Roellig.  The photometry is given in Table 6. 

WISE 1534-10 is of particular interest as it appears to be a cold, extremely metal-poor, halo brown dwarf \citep{Kirkpatrick_2021b, Faherty_2025}.   Searches for distant brown dwarfs may find metal-poor members of the thick disk and halo \citep[e.g.][]{Imig_2023} and so this object, although highly unusual compared to the local brown dwarf sample, may be relevant for deep searches.

\begin{deluxetable*}{lccccc}[b]
\tablecaption{New Synthesized {\it JWST} Photometry, Vega Magnitudes}
\tablehead{
\colhead{Name}  & \colhead{F115W}  & \colhead{F150W}  & 
\colhead{F277W}  & \colhead{F410M}  & \colhead{F444W} 
}
\startdata
WISE J085510.83-071442.5 & 25.66 & 24.00 & 21.63 & 14.69 & 14.08 \\
WISEA J153429.75-104303.3 & 25.40 & 22.58 & 19.70 & 15.81 & 15.78 \\
WISE J182831.08+265037.7 & 23.30 & 23.12 & 20.37 & 14.58 & 14.45 \\
\enddata
\tablecomments{The uncertainty is dominated by the uncertainty in the  
absolute flux calibration of the {\it JWST} instruments, 
which is currently estimated to be 3\% or 0.03 magnitudes.
}
\end{deluxetable*}

\begin{deluxetable*}{lccccc}
\tablecaption{{\it JWST} Colors from \citet{Beiler_2024}  and This Work}
\tablehead{
\colhead{}  & \colhead{$T_{\rm eff}$}  &
\colhead{F277W $-$}  & \colhead{F115W $-$}  & \colhead{F410M $-$}  & \colhead{F115W $-$}   \\
\colhead{Name}  & \colhead{K}  &
\colhead{F444W}  & \colhead{F150W}  & \colhead{F444W}  & \colhead{F277W} 
}
\startdata
WISE J024714.52+372523.5 & 627$^{+35}_{-40}$  & 3.93 & -0.12 & 0.05 & 0.10 \\
WISEP J031325.96+780744.2 & 560$^{+29}_{-31}$ & 4.15   & -0.22  &  0.13   &  0.46  \\
WISE J035934.06-540154.6 & 443$^{+23}_{-19}$ & 4.70 &  -0.29  & 0.12  & -0.32   \\
WISE J043052.92+463331.6 & 495$^{+28}_{-23}$ & 4.20 &  -0.09  &  0.01   & 0.70   \\
WISE J053516.80-750024.9 & 496$^{+28}_{-23}$* & 5.46 &  -0.02    &  0.32 & 2.16   \\
WISE J073444.02-715744.0 & 466$^{+28}_{-23}$ & 4.70 &  -0.36    & 0.17  &  0.73  \\
WISE J082507.35+280548.5 & 387$\pm 15$ & 5.40 & -0.70 & 0.33  &  2.47 \\
WISE J085510.83-071442.5 & 276$\pm 9$  &  7.55 &  1.66  & 0.61 & 4.03 \\
ULAS J102940.52+093514.6 & 705$^{+39}_{-49}$ & 3.21  &  -0.16 &  -0.09 & 0.20   \\
CWISEP J104756.81+545741.6 & 395$^{+23}_{-21}$ & 4.89 & -0.65  & 0.10  &  0.24 \\
WISE J120604.38+840110.6 & 472$^{+26}_{-30}$ &  4.34  &  -0.51  &  0.05 & 1.16   \\
SDSSp J134646.45-003150.4 & 927$^{+58}_{-87}$ &  2.54 &  -0.10 &  -0.29 &  0.17  \\
WISE J140518.39+553421.3 & 392$^{+16}_{-15}$ & 5.83 & -0.35 &  0.31 &  1.30 \\
CWISEP J144606.62-231717.8 & 351$^{+16}_{-13}$  & 5.44 & -0.66  &  0.37 & 1.65 \\
WISE J150115.92-400418.4 & 836$^{+49}_{-68}$ & 2.70 &  -0.18 &  -0.20  &  -0.03  \\
WISEA J153429.75-104303.3 & 502$\pm 6$** & 3.92 & 2.82 &  0.13  &   5.70 \\
WISE J154151.65-225024.9 & 411$^{+18}_{-17}$  & 5.24  & -0.67  &  0.34 &  1.82 \\
SDSSp J162414.37+002915.6 & 922$^{+53}_{-78}$ & 2.80  &  -0.12 &  -0.15 &  0.00  \\
WISE J182831.08+265037.7 & 387$\pm 22$*** & 5.92 & 0.18 & 0.13 & 2.93 \\
WISE J195905.65-333833.5 & 750$^{+42}_{-56}$  & 3.25   &  -0.16  & -0.13  & 0.19   \\
WISE J205628.91+145953.2 & 481$^{+26}_{-20}$  & 4.86  &  -0.45 &  0.14   & 1.15   \\
WISE J210200.15-442919.5 & 549$^{+29}_{-30}$  &  4.17   &  -0.23 &  0.10   &  0.37  \\
WISEA J215949.54-480855.2 & 532$^{+30}_{-28}$ & 4.55    &  -0.15 &  0.06   & 0.26   \\
WISE J220905.73+271143.9 & 355$^{+13}_{-11}$  & 6.36 &  \nodata &  0.34 &  \nodata \\
WISEA J235402.79+024014.1 & 347$^{+13}_{-11}$   & 6.17  &  -0.48 &  0.28 & 1.65 \\
\enddata
\tablenotetext{*}{This Y dwarf may instead be an unresolved binary made up of 480~K and 340~K dwarfs.}
\tablenotetext{**}{~~\citet{Faherty_2025} determine $T_{\rm eff} = 502 \pm 6$ by a retrieval analysis.}
\tablenotetext{***}{~~~Assuming that the Y dwarf is composed of an identical pair of Y dwarfs.}
\tablecomments{The uncertainty in each color is 
estimated to be 4\% or 0.04 magnitudes, and is
dominated by the uncertainty in the  
absolute flux calibration of the {\it JWST} instruments.}

\end{deluxetable*}

\begin{figure}
\vskip -0.2in
\hskip 0.4in
\includegraphics[angle=0, width = 5.0 in]
{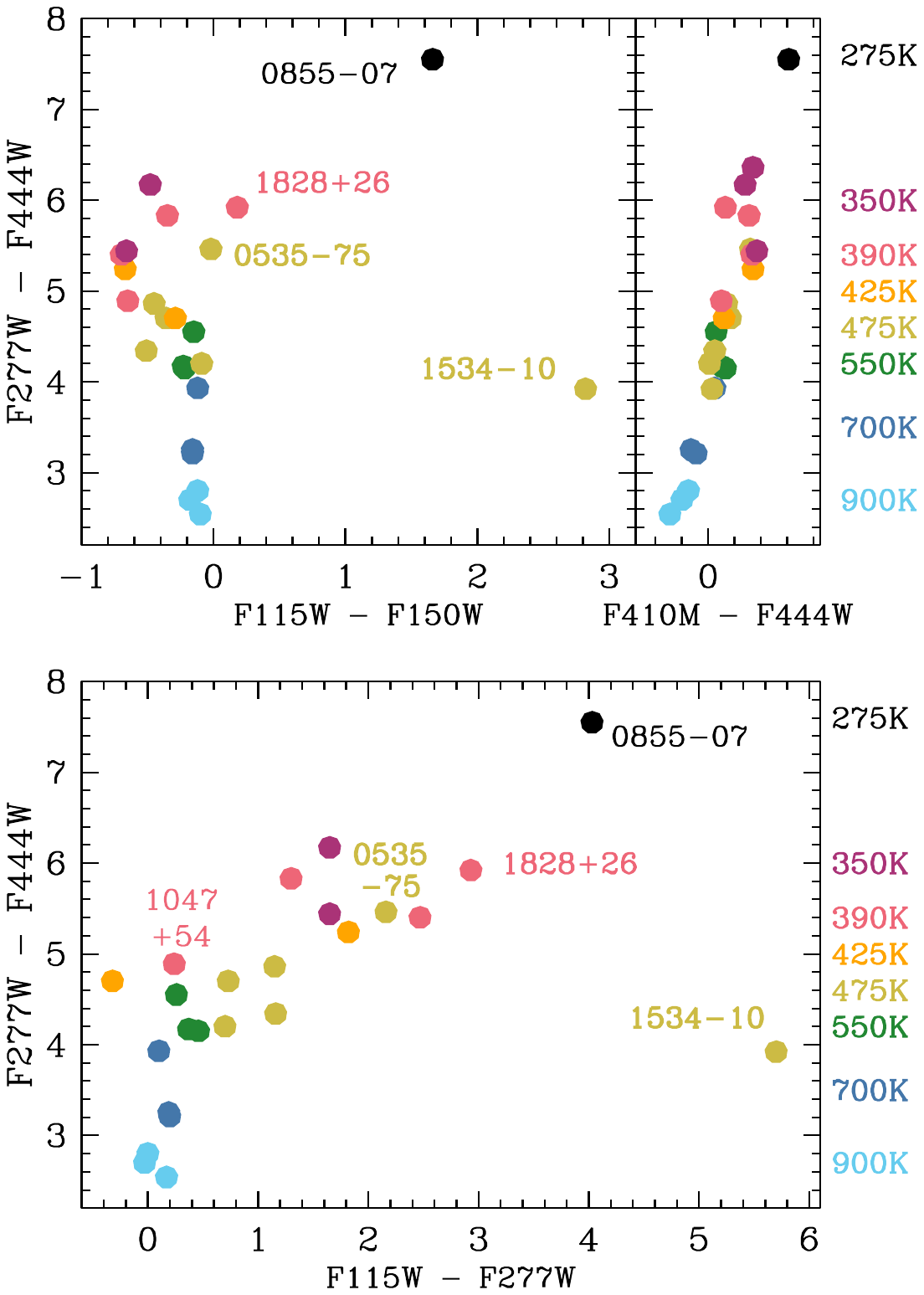}
\vskip -0.3in
\caption{{\it JWST}  colors for the sample in Table 7. Data points are color-coded for $T_{\rm eff}$ as shown in the legends on the right axes. Five brown dwarfs are identified and discussed further in Appendix A.}
\end{figure}

Table 7 lists the {\it JWST} colors used in the \citet{Hainline_2025} search for distant brown dwarfs, where the data is taken from \citet{Beiler_2024} and Table 6.
Figure 7 shows these colors in the \citet{Hainline_2025} color selection diagram (their Figure 1).  The F277W $-$ F444W color increases with decreasing $T_{\rm eff}$ and  Figure 7 suggests that Y dwarfs will have F2777W $-$ F444W $\gtrsim 4.0$.
In F115W $-$ F150W, the cold WISE 0855-07 and very metal-poor high-gravity WISE 1534-10 are significantly redder than the rest of the sample. In F115W $-$ F277W,   WISE 1534-10 is almost 5 magnitudes redder than the more metal-rich and likely younger (and therefore lower gravity, see for example Figure 6) brown dwarfs of similar temperature. The F115W $-$ F277W color also turns to the red for 400~K and cooler Y dwarfs.  As pointed out by \citet{Hainline_2025}, the F410M $-$ F444W color appears insensitive to metallicity and gravity. 

Note that brown dwarfs previously described as unusual have colors somewhat different from objects with similar temperature in Figure 7. WISE 0535-75 is red for its luminosity-based $T_{\rm eff}$; it may be an unresolved binary made up of 480 K and 340 K dwarfs, or it may be metal-poor and high-gravity (Section 3.3).   WISE 1047+54 is blue, and is very likely to be a young, metal-rich, low-gravity dwarf (Section 3.3).  The colors of WISE 1828+26 support the proposed binary nature of this object with $T_{\rm eff} \approx 387$~K, as opposed to it being a single 462~K brown dwarf (Section 2).

\clearpage
\bibliography{Teff_Cold_2026}{}
\bibliographystyle{aasjournal}

\end{document}